\documentclass[amsmath,notitlepage,amssymb,aps,showkeys,floatfix,prd,a4paper,twocolumn,nofootinbib]{revtex4-2}

\usepackage{graphicx}
\usepackage{amsmath}
\usepackage{epstopdf}
\usepackage{amsfonts,amssymb}
\usepackage[utf8]{inputenc}
\usepackage{pstricks}
\usepackage{color}
\usepackage{multirow}
\usepackage{booktabs}
\usepackage{lipsum}

\usepackage[compat=1.0.0]{tikz-feynman}
\usepackage{simplewick}
\usepackage{color, colortbl}
\definecolor{Gray}{gray}{0.9}
\usepackage{float}
\usepackage{tikz-feynman}
\usepackage{feynmp}
\usepackage{forest}

\usepackage{verbatim}    
\usepackage{dcolumn}
\usepackage{bm}
\usepackage{epsfig}
\usepackage{braket}
\usepackage[normalem]{ulem} 
\usepackage{longtable} 

\usepackage{hyperref}
\usepackage{slashed}

\newcommand{\be}{\begin{equation}}
\newcommand{\ee}{\end{equation}}
\newcommand{\ben}{\begin{eqnarray}}
\newcommand{\een}{\end{eqnarray}}

\newcommand{\pslash}{\not{\hbox{\kern-2.3pt $p$}}}
\newcommand{\pdslash}{\not{\hbox{\kern-2pt $\partial$}}}

\begin{document}


\title{Interactions of the deuteron with a hadronic medium}

\author{ L. M. Abreu}
\email{luciano.abreu@ufba.br}
\affiliation{ Instituto de F\'isica, Universidade Federal da Bahia,
Campus Universit\'ario de Ondina, 40170-115, Bahia, Brazil}

\author{R. Higa}
\email{higa@if.usp.br}
\affiliation{Instituto de F\'{\i}sica, Universidade de São Paulo,  
Rua do Mat\~ao 1371,  05508-090
Cidade Universit\'aria, São Paulo, SP, Brazil} 

\author{R. O. Magalhães}
\email{rodrigomagalhaes@ufba.br}
\affiliation{ Instituto de F\'isica, Universidade Federal da Bahia,
Campus Universit\'ario de Ondina, 40170-115, Bahia, Brazil}

\author{F. S. Navarra}
\email{navarra@if.usp.br}
\affiliation{Instituto de F\'{\i}sica, Universidade de São Paulo,  
Rua do Mat\~ao 1371,  05508-090
Cidade Universit\'aria, São Paulo, SP, Brazil} 



\begin{abstract}

We investigate the interactions of the deuteron with light mesons during the hadronic phase in heavy-ion collisions. We treat the deuteron as a weakly bound state and employ the quasi-free approximation to describe the $d\pi$ interaction. The underlying elementary $N\pi$ amplitudes are described by a hybrid effective model, combining the non-resonant background from chiral perturbation theory with  resonant contributions via Breit-Wigner parameterizations. These amplitudes are used to calculate the vacuum and thermally-averaged cross-sections for deuteron dissociation and production, namely, $d + \pi \rightarrow N + N' + \pi$ and the corresponding inverse reaction. We then  use these cross sections in a rate equation to estimate the time evolution of the deuteron multiplicity. For the initial conditions we consider two models: the statistical hadronization model and the coalescence model, where the deuteron is treated as a hadronic molecule. Our findings suggest that the final deuteron yield does not retain a memory of its initial production mechanism.

\end{abstract}
\maketitle

\section{Introduction}
\label{Introduction} 

Understanding deuteron production in high energy nuclear collisions is a challenge for theorists. 
It is amazing that a fragile and weakly-bound object can be formed in such violent collisions.  Indeed, the deuteron ($d$) represents the most fundamental and well-established hadronic molecule: a proton-neutron bound state with small binding energy ($E_B \approx 2.2$ MeV)  and quantum numbers $I(J^P) = 0(1^+)$~\cite{strinati2018bcs}. Its compositeness is confirmed by the Weinberg criterion~\cite{Weinberg:1965zz, matuschek2021nature} and by extensive scattering data. As the simplest nuclear bound state, the deuteron serves as a crucial benchmark for understanding more complex molecular candidates such as the  $X(3872)$.

In recent years we have dedicated  continuous efforts to understand the exotic hadrons 
(X,Y,Z states) \cite{xint-14-16,tcc22}. 
The guideline of our works was to use the hadron gas formed in heavy ion collisions as a filter 
to separate molecular from tetraquark states. The naive expectation that large molecules would be immediately 
destroyed whereas compact tetraquarks would just cross the hadron gas almost undisturbed was not quite confirmed. We still need to know more about the interactions of molecules in the hadron gas. In this context the deuteron emerges 
as a valuable tool. It is undoubtely a molecule and it has been observed in heavy ion collisions.

Recent measurements of deuteron production in HICs~\cite{NA49:2004mrq,ALICE:2015wav,STAR:2019sjh,ALICE:2022veq} have revealed interesting features. For example, in~\cite{bleicher3,SHM1,bleicher4,ALICE:2022veq,toma25} it is argued that both the statistical hadronization model (SHM) and the coalescence model (COM) can describe the data.
This is surprising since these models are quite different. 

In the SHM a particle is ``born-ready" and the only parameters
that determine its yield are the temperature, bariochemical potential and the volume of the system.  The extracted value of temperature is close to the deconfinement transition temperature. This fact supports the picture in which after hadronizing the system ``explodes" and the particles remember the critical temperature because there is no hadron gas phase, no further interactions and no cooling. However this hypothesis seems to be too strong. It seems 
more plausible that after hadronization the system forms a hadron gas which lives for some time.

The duration of the
hadron gas phase is still subject of debate. We can obtain information about the hadron phase by measuring $K^*$ and its suppression. At lower energies, the latest estimates \cite{bleicher1,bleicher2} for the lifetime of the hadronic stage suggest a duration of the order of  4 - 8 fm/c. At higher energies, the duration can even be longer \cite{chiara}. A well-known tool used to extract information on the spatial extent and lifetime of the particle emission source is Hanbury-Brown-Twiss interferometry (HBT). HBT analyses \cite{hbt1,hbt2,hbt3,hbt4} of 
experimental data suggest a lifetime of the hadron gas of $\approx  5 -  15$  fm/c. In view of these estimates, it 
is hard to believe in the sudden explosion scenario, i.e., hadronization immediately followed  by  
kinetic freeze-out. If there is a hadron gas phase, it is very likely that the initial yield of hadrons (formed at the conversion from quark-gluon plasma to a hadron gas) will suffer changes due to rescattering of the considered hadron with the mesons in the hadron gas. In this case, the temperature found in the SHM fits of the data must be interpreted as the temperature from which on the yields no longer change, i.e., the chemical equilibrium temperature. This temperature is lower than the hadronization temperature.

The other well known model of particle production is the COM \cite{ExHIC:2017smd}, which is characterized by the convolution of the density matrix of the constituents of the formed hadron with its Wigner function. This model  has the advantage of carrying information about the intrinsic structure of the system, such as angular momentum and the  type and number of constituents. In this model  deuteron production is achieved by binding nucleons that were produced during hadronization and are close to each other in the phase space. We emphasize that in our approach, as in 
\cite{ExHIC:2017smd}, the coalescence happens during the transition from the quark-gluon plasma to the hadron gas phase and the produced hadron has to live in the hadron gas and interact with other light hadrons. There are other 
coalescence models, such as the one used in \cite{toma25} where the deuteron is formed at the end of 
collision.  In this version of the coalescence one binds nucleons that already decoupled from the hot fireball and 
are close to each other in the phase space. This model does not include the detailed dynamics of how the two nucleons are bound, that were originally considered free.


In this work, we investigate the deuteron as a prototypical hadronic molecule. We calculate its interactions with the hadron medium and the resulting time evolution of its abundance.
One denotes de deuteron ($d$) as a weakly-bound state of one nucleon $N$ with a second one $N'$. The deuteron-pion ($d\pi$) cross section is computed using the quasi-free approximation, where the pion scatters off individual constituent nucleons. The underlying nucleon-pion ($N\pi$) amplitudes combine chiral perturbation theory ($\chi$PT) for the non-resonant background with Breit-Wigner forms for resonant states.
The resulting vacuum cross sections for the $d\pi \leftrightarrow NN'\pi$ processes are thermally averaged to incorporate medium effects. These are then used in a kinetic rate equation to compute the time evolution of the deuteron multiplicity during the hadronic phase.
We initiate this evolution choosing two distinct initial conditions: one provided by the SHM~\cite{Andronic:2017pug} and another by the COM~\cite{ExHIC:2017smd,Yun:2022evm}. This allows us to determine whether the final deuteron yield retains a memory of its initial production mechanism or converges to a universal value governed by in-medium scattering processes.
All the parameters which characterize the hadron gas have already been fixed fitting other data
\cite{ExHIC:2017smd}. The other two inputs which we can test are the initial number of deuterons (and hence the
SHM and COM) and their thermal interaction cross sections. As we demonstrate, the final yield of deuterons is quite insensitive to the assumed initial multiplicity.  This result can be compared with some previous studies, such as Ref.~\cite{Staudenmaier:2021lrg}, in which deuteron formation in HICs was simulated by implementing stochastic multi-particle reactions within a hadronic transport approach. There, it was found that multi-particle interactions substantially reduce the deuteron equilibration time. Another recent result comes from Ref.~\cite{Sun:2022xjr}, which also employs a stochastic approach to show that the deuteron yield remains nearly unchanged during the hadronic expansion phase.

This work is organized as follows. In Sec.~\ref{sec:formalism}, we provide details of the theoretical formalism, including the quasi-free approximation, the effective model for $N\pi$ scattering, and the calculation of vacuum and thermally  averaged cross sections. In Sec.~\ref{Abundance}, we present our results for the time evolution of the deuteron multiplicity. Finally, in Sec.~\ref{Conclusions}, we discuss the implications of our findings and present our concluding remarks.


\section{Interactions of the deuteron with pions}
\label{sec:formalism}

We aim to determine the interactions of the deuteron with light mesons in the hadronic medium formed in heavy-ion collisions. To this end, the following methodology is adopted to calculate the needed thermally-averaged cross sections.

\begin{figure}[!htbp]
\centering
\scalebox{0.8}{\begin{tikzpicture}[thick]
  \begin{feynman}
    \vertex (a);
    \vertex [below left=of a] (i1) {\(N\)};
    \vertex [above left=of a] (i2) {\(\pi\)};
    \vertex [right=3cm of a] (b);
    \vertex [below right=of b] (f1) {\(N\)};
    \vertex [above right=of b] (f2) {\(\pi\)};
    \vertex at ($(a) + (-1.5cm,-2cm)$) (n1) {\(N'\)};
    \vertex at ($(b) + (1.5cm,-2cm)$) (n2) {\(N'\)};
    \node at ($(a) + (1.5,-2.5)$) {(a)};
    
    \diagram* {
      (i1) -- [fermion, very thick] (a) -- [fermion, very thick, edge label'=\(N\)] (b) -- [fermion, very thick] (f1),
      (i2) -- [scalar, very thick] (a),
      (b) -- [scalar, very thick] (f2),
      (n1) -- [fermion, very thick, edge label=\(N'\)] (n2),
    };    
  \end{feynman}
\end{tikzpicture}}
\hspace{0.2cm}
\scalebox{0.8}{\begin{tikzpicture}[thick]
  \begin{feynman}
    \vertex (a);
    \vertex [below left=of a] (i1) {\(N'\)};
    \vertex [above left=of a] (i2) {\(\pi\)};
    \vertex [right=3cm of a] (b);
    \vertex [below right=of b] (f1) {\(N'\)};
    \vertex [above right=of b] (f2) {\(\pi\)};
    \vertex at ($(a) + (-1.5cm,-2cm)$) (n1) {\(N\)};
    \vertex at ($(b) + (1.5cm,-2cm)$) (n2) {\(N\)};
    \node at ($(b) + (-1.5,-2.5)$) {(b)};
    
    \diagram* {
      (i1) -- [fermion, very thick] (a) -- [fermion, very thick, edge label'=\(N'\)] (b) -- [fermion, very thick] (f1),
      (i2) -- [scalar, very thick] (a),
      (b) -- [scalar, very thick] (f2),
      (n1) -- [fermion, very thick, edge label=\(N\)] (n2),
    };
  \end{feynman}
\end{tikzpicture}}

\vspace{0.2cm}

\scalebox{0.8}{\begin{tikzpicture}[thick]
  \begin{feynman}
    \vertex (a);
    \vertex [left=1cm of a] (i1) {\(N\)};
    \vertex [above=of i1] (i2) {\(\pi\)};
    \vertex [right=3.0cm of a] (b);
    \vertex [right=1cm of b] (f1) {\(N\)};
    \vertex [above=of f1] (f2) {\(\pi\)};
    \vertex at ($(a) + (-1.5cm,-1.cm)$) (n1) {\(N'\)};
    \vertex at ($(b) + (1.5cm,-1.cm)$) (n2) {\(N'\)};
    \node at ($(a) + (1.5,-1.5)$) {(c)};    
    \diagram* {
      (i1) -- [fermion, very thick] (a) -- [fermion, very thick, edge label'=\(N\)] (b) -- [fermion, very thick] (f1),
      (i2) -- [scalar, very thick] (b),
      (f2) -- [scalar, very thick] (a),
      (n1) -- [fermion, very thick, edge label=\(N'\)] (n2),
    };    
  \end{feynman}
\end{tikzpicture}}
\hspace{0.2cm}
\scalebox{0.8}{\begin{tikzpicture}[thick]
  \begin{feynman}
    \vertex (a);
    \vertex [left=1cm of a] (i1) {\(N'\)};
    \vertex [above=of i1] (i2) {\(\pi\)};
    \vertex [right=3.0cm of a] (b);
    \vertex [right=1cm of b] (f1) {\(N'\)};
    \vertex [above=of f1] (f2) {\(\pi\)};
    \vertex at ($(a) + (-1.5cm,-1.cm)$) (n1) {\(N\)};
    \vertex at ($(b) + (1.5cm,-1.cm)$) (n2) {\(N\)};
    \node at ($(a) + (1.5,-1.5)$) {(d)};    
    
    \diagram* {
      (i1) -- [fermion, very thick] (a) -- [fermion, very thick, edge label'=\(N'\)] (b) -- [fermion, very thick] (f1),
      (i2) -- [scalar, very thick] (b),
      (f2) -- [scalar, very thick] (a),
      (n1) -- [fermion, very thick, edge label=\(N\)] (n2),
    };
  \end{feynman}
\end{tikzpicture}}

\vspace{0.2cm}

\scalebox{0.8}{\begin{tikzpicture}[thick]
  \begin{feynman}
    \vertex (a);
    \vertex [left=2.5cm of a] (i1) {\(N\)};
    \vertex [above=of i1] (i2) {\(\pi\)};
    \vertex [right=2.5cm of a] (f1) {\(N\)};
    \vertex [above=of f1] (f2) {\(\pi\)};
    \vertex at ($(a) + (-3.cm,-1.cm)$) (n1) {\(N'\)};
    \vertex at ($(a) + (3.cm,-1.cm)$) (n2) {\(N'\)};
    \node at ($(a) + (0,-1.5)$) {(e)};    
    
    \diagram* {
      (i1) -- [fermion, very thick] (a) -- [fermion, very thick] (f1),
      (i2) -- [scalar, very thick] (a) -- [scalar, very thick] (f2),
      (n1) -- [fermion, very thick, edge label=\(N'\)] (n2),
    };
    
  \end{feynman}
\end{tikzpicture}}
\hspace{0.2cm}
\scalebox{0.8}{\begin{tikzpicture}[thick]
  \begin{feynman}
    \vertex (a);
    \vertex [left=2.5cm of a] (i1) {\(N'\)};
    \vertex [above=of i1] (i2) {\(\pi\)};
    \vertex [right=2.5cm of a] (f1) {\(N'\)};
    \vertex [above=of f1] (f2) {\(\pi\)};
    \vertex at ($(a) + (-3cm,-1.cm)$) (n1) {\(N\)};
    \vertex at ($(a) + (3cm,-1.cm)$) (n2) {\(N\)};
    \node at ($(a) + (0,-1.5)$) {(f)};    
    
    \diagram* {
      (i1) -- [fermion, very thick] (a) -- [fermion, very thick] (f1),
      (i2) -- [scalar, very thick] (a) -- [scalar, very thick] (f2),
      (n1) -- [fermion, very thick, edge label=\(N\)] (n2),
    };
  \end{feynman}
\end{tikzpicture}}

\caption{Tree-level diagrams contributing to the $d\pi$ reactions, according to the quasi-free approximation approach~\cite{quase_livre}, in which they are considered as the sum over the processes where the pion scatters from one of the constituent nucleons, with the complementary nucleon acting as a spectator in each case. Diagrams (a) and (b) stand for processes involving the $s$-channel $N^{(\prime)}\pi \to N^{(\prime)}\pi$ scattering, (c) and (d) for processes involving the $u$-channel $N^{(\prime)}\pi \to N^{(\prime)}\pi$ scattering, and (e) and (f) for processes involving the contact-type $N^{(\prime)}\pi \to N^{(\prime)}\pi$ scattering.}
\label{fig:diagramas}
\end{figure}
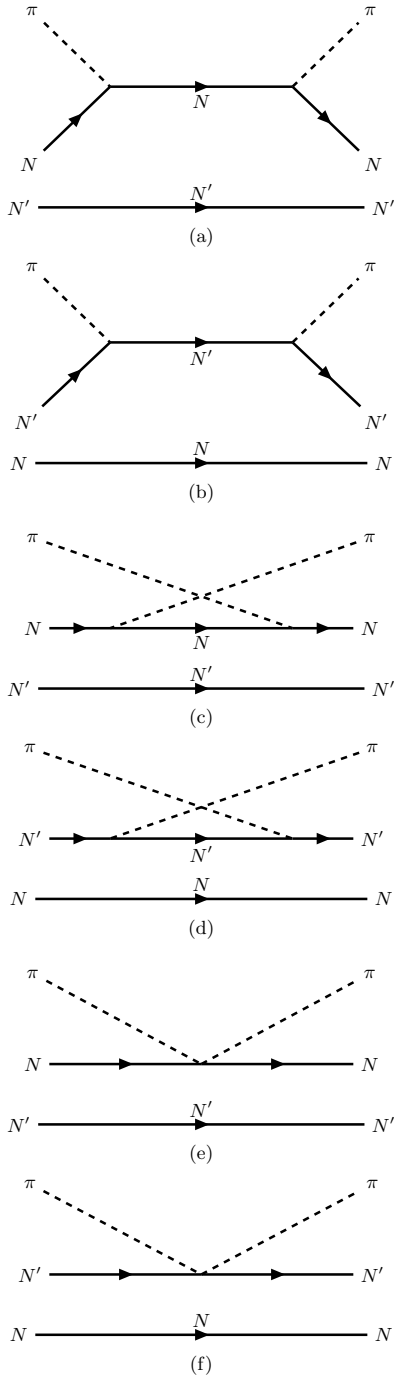

\begin{itemize}

    \item Based on our prior analysis of other states (see, for example, Refs~\cite{xint-14-16,jpsi-18,dstar22,tcc22,zczcs,chi4274,psi2s}), we have found that pions (the most abundant constituents) provide the most important contributions for an initial description of the hadronic environment. Consequently, we restrict our analysis to interactions between  deuterons and pions.

    \item To evaluate the cross sections of the $d$ dissociation process $d + \pi \rightarrow N + N^\prime + \pi$ and its inverse reaction we employ  the quasi-free approximation approach~\cite{Chew1952, Kerman1959} (see also \cite{quase_livre}). Consequently, the $d\pi$ interaction is modeled by adding the processes in which the pion scatters off one of the constituent nucleons while the other constituent nucleon acts as a spectator, as depicted in Fig.~\ref{fig:diagramas}.

    \item The scattering amplitudes for processes involving nucleons ---protons $(p)$ or neutrons $(n)$--- and pions are described within a hybrid effective model inspired by the isobar model. This approach combines contributions from $\chi$PT~\cite{Scherer:2012xha}, which provides the non-resonant background, with explicit Breit-Wigner parameterizations for the relevant resonant states.

    \item The vacuum $d\pi$ cross sections are used as input to compute the thermally averaged cross sections. This is essential for modeling the deuteron behavior in a heavy-ion collision environment, where its interactions with light hadrons are influenced by the hot medium. The thermal average incorporates the distribution of collision energies at a given temperature $T$ of the surrounding medium.

\end{itemize}

This procedure is detailed in what follows.

\subsection{$N - \pi$ interactions}\label{subsecA}

The lowest-order Born diagrams that contribute to the scattering amplitude  $N^{(\prime)}\pi \to N^{(\prime)}\pi$ are shown in Fig.~\ref{fig:diagramas}. 
The corresponding amplitude is obtained in the framework of chiral effective field theory as pedagogically spelled out in Ref.~\cite{Scherer:2012xha}. 
For completeness, we briefly summarize the essential aspects here. The interaction vertices at leading order in the chiral expansion are derived from the effective Lagrangian
\begin{align} \label{eq: lagrangiana}
	\mathcal{L}_{int}
					= &-\dfrac12\frac{g_A}{F_\pi}\bar{N}\gamma^\mu \gamma_5 \ \partial_\mu\vec{\pi} \cdot \vec{\tau} \ N
					\nonumber \\& -\dfrac1{4F^2}\bar{N}\gamma^\mu\ \vec{\pi}\times\partial_\mu\vec{\pi}\cdot\vec{\tau}\ N,
	\end{align}
where $\vec{\pi}$ denotes the pion isospin triplet and $\vec{\tau}$ the Pauli matrices in isospin space,  $N = (p, n)$ is the nucleon isospin doublet, $g_A = 1.267$ is the coupling constant, and $F_\pi = 92.4 \, \text{MeV}$ is the pion decay constant.
The first term in Eq.~\eqref{eq: lagrangiana} corresponds to the pion-nucleon-nucleon coupling while the second term is the Weinberg-Tomozawa contact interaction involving two pions and two nucleons. 

We write the invariant amplitude $\mathcal{M} = iT$  for the processes $\pi_a(q) + N(p) \to \pi_b(q^\prime) + N(p^\prime)$, obtained from Eq.~\eqref{eq: lagrangiana}, by means of the parametrization~\cite{Scherer:2012xha} 
\begin{align}
		T_{ab}(p, q; p^\prime, q^\prime)
		&=\delta_{ab}T^+(p, q; p^\prime, q^\prime)  \nonumber \\ &-i\epsilon_{abc}\tau_cT^-(p, q; p^\prime, q^\prime), 
	\label{ampl_t1}
    \end{align}
where 	
	\begin{align}\label{TPlusTMinus}
		T^+  = & \dfrac{g_A^2}{4F_\pi^2}\bar{u}(p^\prime) \left\{ \left[ 2m_N+\dfrac{1}{2}\left(\slashed{q}+\slashed{q}^\prime \right)\left(-1-\dfrac{2m_N}{\nu-\nu_B}\right)\right] \right. \nonumber \\ & + \left. \left[ 2m_N+\dfrac{1}{2}\left(\slashed{q}+\slashed{q}^\prime \right)\left(1-\dfrac{2m_N}{\nu+\nu_B}\right)\right] \right\}  u(p) , \nonumber \\ 
		T^-= &\dfrac{g_A^2}{4F_\pi^2}\bar{u}(p^\prime) \left\{ \left[ 2m_N+\dfrac{1}{2}\left(\slashed{q}+\slashed{q}^\prime \right)\left(-1-\dfrac{2m_N}{\nu-\nu_B}\right)\right] \right. \nonumber \\  & + \left. \left[ 2m_N+\dfrac{1}{2}\left(\slashed{q}+\slashed{q}^\prime \right)\left(1-\dfrac{2m_N}{\nu+\nu_B}\right)\right] \right\}  u(p)  \nonumber \\ &+ \bar{u}(p^\prime)\dfrac{(\slashed{q}+\slashed{q}^\prime)}{4F_{\pi}^2}u(p),
	\end{align}
and 
\begin{eqnarray}
		\nu = \dfrac{s-u}{4m_N}, \ \ \nu_B = \dfrac{t-2m_\pi^2}{4m_N},
	\end{eqnarray}
with $s$, $u$, and $t$ being the Mandelstam variables, and $m_\pi, m_N$ the pion and nucleon masses, respectively. 
The first terms between brackets in $T^{\pm}$ given in Eq.~\eqref{TPlusTMinus} correspond to the $s$-channel while the second ones are associated with the $u$-channel. The last term in $T^-$ comes from the Weinberg-Tomozawa contact term.

To express our scattering amplitudes in terms of the channels $T^+$ and $T^-$ it is more convenient to project them into a well-defined isospin basis. We consider the $N\pi$ system with the nucleon isodoublet and the pion isotriplet. Decomposing this system into its total isospin components one writes the amplitudes for the isospin-$\frac{1}{2}$ and isospin-$\frac{3}{2}$ channels as follows:
\begin{align}
|p\pi^+\rangle & \equiv  \Bigl|\frac{3}{2},\frac{3}{2}\Bigr\rangle, \\[1ex]
|p\pi^0\rangle & \equiv  \sqrt{\frac{2}{3}}\,\Bigl|\frac{3}{2},\frac{1}{2}\Bigr\rangle + \sqrt{\frac{1}{3}}\,\Bigl|\frac{1}{2},\frac{1}{2}\Bigr\rangle, \\[1ex]
|p\pi^-\rangle & \equiv  \sqrt{\frac{1}{3}}\,\Bigl|\frac{3}{2},-\frac{1}{2}\Bigr\rangle + \sqrt{\frac{2}{3}}\,\Bigl|\frac{1}{2},-\frac{1}{2}\Bigr\rangle, \\[1ex]
|n\pi^+\rangle & \equiv  \sqrt{\frac{1}{3}}\,\Bigl|\frac{3}{2},\frac{1}{2}\Bigr\rangle - \sqrt{\frac{2}{3}}\,\Bigl|\frac{1}{2},\frac{1}{2}\Bigr\rangle, \\[1ex]
|n\pi^0\rangle & \equiv  \sqrt{\frac{2}{3}}\,\Bigl|\frac{3}{2},-\frac{1}{2}\Bigr\rangle - \sqrt{\frac{1}{3}}\,\Bigl|\frac{1}{2},-\frac{1}{2}\Bigr\rangle, \\[1ex]
|n\pi^-\rangle & \equiv  \Bigl|\frac{3}{2},-\frac{3}{2}\Bigr\rangle, 
\label{Npiprojections}
\end{align}
where the notation $|I, I_3\rangle$ ($I$ being the total isospin and $I_3$ its projection over the third axis) has been used. 
Using these relations in Eq.~(\ref{ampl_t1}), we obtain the  standard isospin-projected scattering amplitudes written as
\begin{align}
T^{\left( \frac{3}{2} \right)} &= T^+ - T^-, \nonumber \\
T^{\left( \frac{1}{2} \right)} &= T^+ + 2T^-.
\label{ampl_t2}
\end{align}

\subsection{$d - \pi$ interactions}\label{subsecB}

The structure of the deuteron has a long history of scientific investigation and is quite well established. It consists of a proton and a neutron weakly bound by $\sim 2 \ \mathrm{MeV}$, with total spin $S=1$ and positive parity. It is mostly an $s$-wave, with a small $d$-wave component due to the tensor force. Its wave function is symmetric both in position and spin spaces, and anti-symmetric in isospin space ($I=0$), as required by the Pauli exclusion principle. One denotes the deuteron state in terms of two nucleons as $|d\rangle=|NN'\rangle$. In isospin space, it can be written as
\begin{eqnarray}
\ket{d} & \equiv & \ket{I=0, I_3=0} \nonumber \\
    &  \equiv  & \frac{1}{\sqrt{2}}(\ket{pn} - \ket{np}).
\label{d-decomp}
\end{eqnarray}

One evaluates the scattering amplitude for the dissociation process $d+\pi \to N + N' + \pi$ within the quasi-free approximation. 
This approximation, also known as the impulse approximation (or even single-scattering approximation)~\cite{Chew1952, Kerman1959,quase_livre}, is often used in nuclear physics and provides a framework for describing interactions where an incident particle scatters primarily from a single constituent within a composite target, while the remaining constituents act as spectators. It is valid when (i) the incident particle energy is sufficiently high so that the binding energy of the system can be neglected during the scattering process, and (ii) the momentum transfer is large enough that the incident particle never interacts strongly with more than one constituent of the system. The present case of $d\pi$ scattering processes, where the pion wavelength is short compared to the deuteron size,  appears particularly  appropriate for this approximation (see for example the discussion in Ref.~\cite{Scheibl1999}). 

In our quasi-free approach, the dissociation process comprises the pion-nucleon scattering amplitude evaluated between deuteron states, c.f. Eq.~(\ref{d-decomp}). Given the isospin-zero content of the latter, the only nonzero contributions come from processes in which the struck nucleon does not change its isospin. Thus, the dissociation amplitude can only be proportional to $T^{+}$, the statistical average over the $I=3/2$ and $I=1/2$ channels. 
At the cross-section level, a more detailed calculation leads to 
\begin{equation}
|\langle d\pi | \hat{T}_{d\pi \to NN'\pi} | d\pi \rangle|^2 = 2|T^+|^2,
\label{sq_amplpiplus2}
\end{equation}
where the factor of 2 in the r.h.s. reflects the incoherent sum of each nucleon in the deuteron being hit by the incident pion. 

The above isospin-symmetric cross section alows one to obtain the total $d\pi$ cross section in terms of $d\pi^{+}$, namely, $\sigma_{d\pi}^{\mathrm{total}} = 3 \times \sigma_{d\pi^+}$. We are aware that the individual cross sections $\sigma_{d\pi^+}$, $\sigma_{d\pi^0}$, and $\sigma_{d\pi^-}$ are not identical (see, for instance, Ref.~\cite{Workman:2012hx,Mathieu:2015gxa,Kong:2018bst}). However, we assume that these differences do not play a significant role in the thermally-averaged result and are therefore not expected to alter the qualitative conclusions of this study.

One knows that $\chi$PT provides a robust and model-independent framework for pionic processes at low energies. However, the formalism is constrained by the spontaneous chiral symmetry energy gap of $\Lambda_{\chi}\sim 1 \ \mathrm{GeV}$~\cite{Scherer:2012xha,Bedaque:2002mn,Epelbaum:2008ga}. Its radius of convergence is typically of the order of 1-2 times the pion mass, $200$-$300 \ \mathrm{MeV}$, above threshold. To model the above expressions as background processes at higher energies, one takes into account the finite size of the hadrons and suppress the artificial growth of the cross sections at large momenta by multiplying each vertex of the amplitudes with a monopole-like form factor,
\begin{equation}
F(\vec{q})=\frac{\Lambda^2}{\Lambda^2+\vec{q}^2}.
\label{eq:ffactor}
\end{equation}
Here, $\vec{q}$ is the transferred three-momentum in the corresponding channel. In the case of the contact contributions, we multiply them by $F^2$ with $\vec{q}^2 = |\vec{p}_i|^2 + |\vec{p}_f|^2$, where $\lvert \vec{p}_{i} \rvert$ and $\lvert \vec{p}_{f} \rvert$ are the magnitudes of the three-momenta of the initial and final particles, respectively, in the CM frame.
The cutoff scale $\Lambda\sim 3 \ \mathrm{GeV}\gtrsim \Lambda_{\chi}$ allows for the desired high-energy extension while guarantees that chiral predictions remain intact at low energies.
This type of form factor has been extensively employed in literature, and a  discussion on its  role is found in Ref.~\cite{tcc22}.




\subsection{Inclusion of resonant states}
\label{formfactor}

\begin{figure}[!htbp]
		\center
			\includegraphics[{width=0.95\linewidth}]{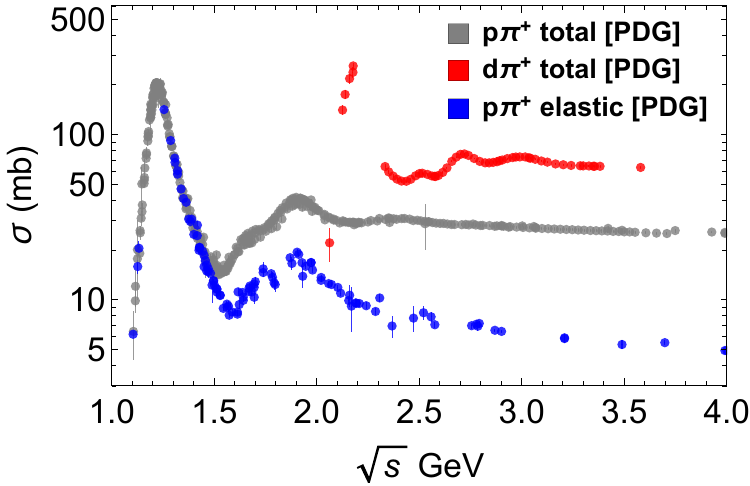}
			\caption{Experimental cross sections for the $p\pi^+$ elastic, $p\pi^+$ total and $d\pi^+$ total scattering processes extracted from Ref~\cite{ParticleDataGroup:2024cfk}.}
			\label{fig:dados_total}
\end{figure}

As mentioned before, chiral effective field theory for nuclear forces are limited to relative energies well below the breakdown scale $\Lambda_{\chi}\sim 1\mathrm{GeV}$~\cite{Bedaque:2002mn,Epelbaum:2008ga}. The interactions built in previous subsections \ref{subsecA} and \ref{subsecB}, valid for energies $\sim 300 \ \mathrm{MeV}$ above the nucleon mass, can be extended to higher energies with the help of a form factor like Eq.~(\ref{eq:ffactor}). However, they cannot account for the rich resonant phenomena of the nucleon spectrum. For instance, the elastic scattering process $p\pi^+ \to p\pi^+$ exhibit peaks in the region below $\sqrt{s}= 2\ \mathrm{GeV}$ related to the prominent resonances  $\Delta(1232)$ and $\Delta(1920)$ in the $P_{33}$ partial wave. 
Their signatures can be seen in Fig.~\ref{fig:dados_total}, where the experimental curves of the cross sections are shown for completeness. 
To extend our description of the nucleon-pion cross sections to higher energies and reproduce the resonant structures observed experimentally, we adopt the framework of the isobar model~\cite{ParticleDataGroup:2024cfk}. This approach describes particle production cross-sections by decomposing them into a sum of intermediate resonant states on top of a smooth background, which in this work, is given by the interactions in \ref{subsecA}, \ref{subsecB}. Each resonance contributes by means of a Breit-Wigner mass distribution and specific angular dependencies based on its spin. 

Accordingly, the resulting amplitude for the $p\pi^+$ channel is given by a combination of the nonresonant contribution given by Eq.~(\ref{ampl_t2}) and the resonant contributions $T_{\mathrm{BW}}$:
\begin{equation}\label{eq:ampPPiPlus}
	\langle p\pi^+ | \hat{T} | p\pi^+ \rangle = T^+ - T^- + \sum _a T_{\mathrm{BW}}^{(a)} ,
\end{equation}
where $T_{\mathrm{BW}}^{a}$ ($a=\Delta(1232) , \Delta(1920)$) is the full Breit-Wigner amplitude for a resonance produced in channel $a$ embedded into the hadronic current, which in the case of a resonance decaying into a nucleon and a pion reads
\begin{equation}\label{eq:BW_current}
	T_{\mathrm{BW}}^{(a)} = \bar{u}(p^\prime) (\slashed{q} + \slashed{q}^\prime) \, \mathcal{A}_a(s) \, u(p).
\end{equation}
The function $ \mathcal{A}_a(s)$ is the standard Breit-Wigner amplitude parameterized as a function of the center-of-mass (CM) energy $s$~\cite{ParticleDataGroup:2024cfk}:
\begin{equation}\label{eq:BW1}
	\mathcal{A}_a(s) = \frac{\mathcal{N}_a(s)}{M^2_{\mathrm{BW}} - s - i M_{\mathrm{BW}} \Gamma(s)},
\end{equation}
where $M_{\mathrm{BW}}$ is the nominal resonance mass. The energy-dependent total width $\Gamma(s)$ accounts for all decay channels $b$:
\begin{equation}\label{eq:Gamma}
	\Gamma(s) = \frac{1}{M_{\mathrm{BW}}} \sum_b g_b^2 \, \rho_b(s) \, n_b^2(s),
\end{equation}
and the numerator $\mathcal{N}_a(s)$ incorporates process-dependent kinematic factors and couplings:
\begin{equation}\label{eq:Nas}
	\mathcal{N}_a(s) = \alpha \, g_a \, n_a(s).
\end{equation}

In these expressions, the index $b$ runs over all decay channels, $g_b$ are channel-specific coupling constants, and $\rho_b(s)$ is the two-body phase space factor:
\begin{equation}
	\rho_b(s) = \frac{1}{16\pi} \frac{2|\vec{q}_b|}{\sqrt{s}},
\end{equation}
where the breakup momentum $q_b$ is given by
\begin{equation}\label{eq:trimomento}
	q_b = \frac{\sqrt{\lambda(s, m_{1,b}^2, m_{2,b}^2)}}{2\sqrt{s}}.
\end{equation}

The factor $n_a(s)$ ensures the correct angular momentum barrier at threshold and is defined as
\begin{equation}\label{eq:n_a}
	n_a(s) = \left( \frac{q_a}{q_0} \right)^{\! l_a} F_{l_a}\!\left( \frac{q_a}{q_0} \right),
\end{equation}
where $l_a$ is the orbital angular momentum in channel $a$. The phenomenological form factor $F_{l_a}(z)$ (with $z = q/q_0$) compensates for the unphysical growth of the centrifugal factor $(q_a)^{l_a}$ at high momenta and is essential to satisfy positivity constraints that require the dressed propagator to vanish no faster than $1/s$ asymptotically. Commonly, Blatt–Weisskopf form factors are used:
\begin{align}
	F_0^2(z) &= 1, \nonumber \\
	F_1^2(z) &= \frac{1}{1 + z^2}, \nonumber \\
	F_2^2(z) &= \frac{1}{9 + 3z^2 + z^4}.
\label{BWFF}
\end{align}

To connect the model to the observed resonance properties, the coupling $g_b$ for a channel can be related to the partial width at the pole ($s = M_{\mathrm{BW}}^2$). If the total width $\Gamma(M_{\mathrm{BW}})$ is known, the coupling is given by:
\begin{equation}\label{eq:acoplamento_gb}
	g_b = \sqrt{ \frac{ \Gamma(M_{\mathrm{BW}}) \cdot M_{\mathrm{BW}} }{ \rho_b(M_{\mathrm{BW}}) \cdot n_b^2(M_{\mathrm{BW}}) } }.
\end{equation}
For the elastic processes considered here, we assume $g_a = g_b$.

Now we move on to the $ d \pi^+$ process. We employ the quasi-free approximation discussed above. In particular, the scattering amplitude for the $d\pi^+$ channel is given by a combination of the nonresonant contribution from Eq.~(\ref{sq_amplpiplus2}) and the resonant contributions,
\begin{equation}
    	\langle d\pi^+ | \hat{T} | d\pi^+ \rangle = T^+ + \sum _a T_{\mathrm{BW}}^{(a)} ,
\label{ampl_dpiplustotal2}
\end{equation}
where $T_{\mathrm{BW}}^{(a)}$ is the full Breit-Wigner amplitude defined previously, parameterized for the relevant resonances.

\subsection{Vacuum cross sections}
\label{CrossSection}

With the resonant Breit-Wigner peaks and the nonresonant background amplitudes determined in the previous sections  we are able to calculate the relevant cross sections. In the center-of-mass (CM) frame the total cross section is given by
\begin{equation} \label{eq:seccao-choque-total}
\sigma \;=\; 
\frac{1}{64\,\pi^2\,s \,g_i} \,
\frac{\lvert \vec{p}_{f}\rvert}{\lvert \vec{p}_{i}\rvert}
\int d\Omega \;\overline{\sum} \lvert \mathcal{M}\rvert^2 \,,
\end{equation}
where $g_{i} $ is the spin-isospin degeneracy factor of the particles in the initial state, $s$ is the squared center-of-mass energy,  and $\overline{\sum}$ denotes the sum over spins and isospins of the initial and final states.

In the calculations of the cross sections, the isospin-averaged masses reported in Ref.~\cite{ParticleDataGroup:2024cfk} have been used. Besides, we remember the values of the constants $g_A = 1.267 $ and $F_{\pi} = 92.4 \, \text{MeV}$. 

\begin{table}[!htbp]
\centering
\caption{Parameters for the Breit-Wigner amplitudes obtained from the fit to the PDG cross-section data~\cite{ParticleDataGroup:2024cfk}.  }
\label{tab:resonance_parameters}
\begin{tabular}{lccc}
\hline
\hline
Parameter & $\Delta(1232)$ & $\Delta(1920)$  &  $\mathcal{D}(2180)$ \\
\hline
$M_{\mathrm{BW}}$ (GeV) & 1.250 & 1.830 & 2.180 \\
$\Gamma(M_{\mathrm{BW}})$ (GeV) & 0.070 & 0.240 & 0.070 \\
$q_0$ (GeV) & 0.24 & 0.67 & 0.9 \\
$g_a = g_b$ & 5.27 & 10.58 & 6.5\\
$l$ & 1 & 1 & 1 \\
$\alpha$ & $9.0 \pm 1.2$ & $2.75 \pm 0.25$  & $7.0 \pm 0.5$ \\
\hline
\hline
\end{tabular}
\end{table}

We start by presenting the results of the cross section for the $p\pi^+$ elastic channel, obtained using the amplitude given in Eq.~(\ref{eq:ampPPiPlus}) in (\ref{eq:seccao-choque-total}). Fitted parameters for the Breit-Wigner amplitudes of the lowest-energy $P_{33}$ resonances, $\Delta(1232)$ and $\Delta(1920)$, are given in Table~\ref{tab:resonance_parameters}. 
The cutoff parameter for the non-resonant amplitude in Eq.~(\ref{eq:ampPPiPlus}) is $\Lambda = 3.1 \pm 0.3\,\mathrm{GeV}$. In 
addition, to take into account the uncertainties inherent in the parameters and the model, the results are presented in terms of bands associated 
with the smallest and largest possible values of $\Lambda$ and $\alpha$.

\begin{figure}[!htbp]
\center
\includegraphics[{width=0.95\linewidth}]{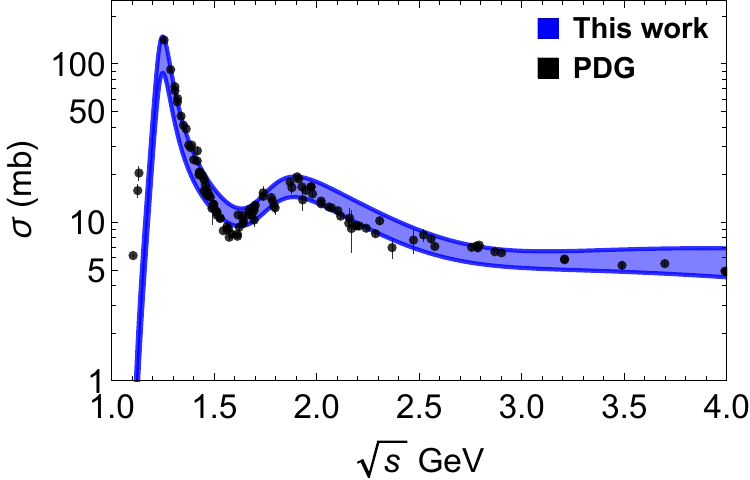}
\caption{Cross section of the process $p\pi^+ \to p\pi^+$ as a function of the CM energy $\sqrt{s}$.}
\label{fig:PPIPLUSADJUST}
\end{figure}	

Our results for the cross section of the elastic $p\pi^+$ scattering channel are displayed in Fig.~\ref{fig:PPIPLUSADJUST}. The theoretical curve shows a good fit to the experimental data across the considered energy range. Notably, the model accurately captures the position, magnitude and width of the resonant peaks, particularly the dominant $\Delta(1232)$ and $\Delta(1920)$ resonances at smaller energies. The successful reproduction of these key features gives strong support to the adopted formalism.

Let us move on to the $ d  \pi^+$ process. The results of the cross section have been obtained replacing the amplitude given in Eq.~(\ref{eq:ampPPiPlus}) by Eq.~(\ref{ampl_dpiplustotal2}), and using it in (\ref{eq:seccao-choque-total}). The  fitted parameters of the Breit-Wigner amplitudes of the lowest-energy peak present in the experimental cross section, denoted as $\mathcal{D}(2180)$, are given in Table~\ref{tab:resonance_parameters}. 
The cutoff parameter for the non-resonant amplitude in Eq.~(\ref{ampl_dpiplustotal2}) is $\Lambda = 2.1 \pm 0.1\,\mathrm{GeV}$.  

\begin{figure}[!htbp]
		\center
			\includegraphics[{width=0.95\linewidth}]{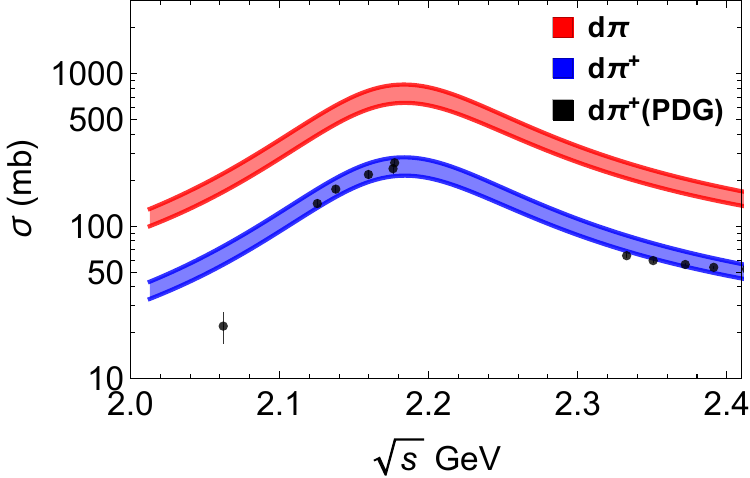}
			\caption{Vacuum cross section of the process  $d\pi^+ \to NN^\prime\pi^+$, and the total $d\pi \to NN^\prime\pi$ as a function of $\sqrt{s}$ to be used in the thermally-averaged cross section. }
			\label{fig:CrSec3d}
\end{figure}

The computed cross section for the $d \pi^+$ reaction is presented in Fig.~\ref{fig:CrSec3d}. The model provides a good description of the experimental data in the low-energy region. This fair reproduction of both the magnitude and energy dependence of the cross section at low energies validates the model  parameterization.

However, deviations from the data emerge at higher energies. These discrepancies highlight the growing contribution of inelastic channels and more complex resonant structures not included in the current theoretical framework.

This limitation is not prejudicial to the primary objective of this work. As detailed in the following subsection, the key quantity for calculating the final deuteron multiplicity is the thermally-averaged cross section. This quantity possesses a crucial property ---it naturally suppresses contributions at very low energies, close to the reaction threshold. On the other hand, the hadron gas considered here exists only at temperatures smaller than $ \approx 180$ MeV. 
Above this value the hadron gas becomes a quark-gluon plasma. This upper limit in the temperature introduces an 
upper limit in  the $ d \pi^+$ collision energy.  
Consequently, the higher-energy deviations observed in the vacuum cross section have a small impact on the thermally  averaged value. 


\subsection{Thermally-averaged cross sections}
\label{ThermalAverage}

In the hadron gas formed in  a heavy-ion collision,  deuterons can undergo interactions with light hadrons. To account for this, it is essential to evaluate the thermal averages of the relevant cross sections, as the collision energy is intrinsically linked to the temperature of the medium.
The thermally averaged cross section is defined as the convolution of the vacuum cross section with the momentum distributions of the colliding particles in the thermal bath. For a generic reaction $a b \to c d$, it is given by~\cite{Koch:1986}
\begin{align}\label{eq: medias_termicas}
\langle \sigma_{ab \to cd}v_{ab} \rangle 
=& \frac{\displaystyle \int d^3\mathbf{p}_a \,d^3\mathbf{p}_b \; f_a(\mathbf{p}_a)\,f_b(\mathbf{p}_b)\,\sigma_{ab \to cd}\,v_{ab}}
       {\displaystyle \int d^3\mathbf{p}_a \,d^3\mathbf{p}_b \; f_a(\mathbf{p}_a)\,f_b(\mathbf{p}_b)} 
\nonumber \\[1em]
=& \frac{1}{4\,\alpha_a^2\,K_2(\alpha_a)\;\alpha_b^2\,K_2(\alpha_b)} 
    \nonumber \\ & \times \int_{z_{0}}^{\infty} \!dz\,K_1(z)\,
    \sigma\bigl(s = z^2 T^2\bigr)\,
    \nonumber \\ & \times \bigl[z^2 - (\alpha_a + \alpha_b)^2\bigr]\,
    \bigl[z^2 - (\alpha_a - \alpha_b)^2\bigr],
\end{align}
where $v_{ab}$ represents the initial relative velocity of the two interacting particles $a$ and $b$, $f_i(\mathbf{p}_i)$ is the momentum distribution 
$\alpha_i = m_i/T$, where $T$ is the temperature, $z_0 = \max(\alpha_a + \alpha_b,\;\alpha_c + \alpha_d)$, and $K_1$ and $K_2$ are the modified Bessel functions.

\begin{figure}[!htbp]
            \center
			\includegraphics[{width=0.95\linewidth}]{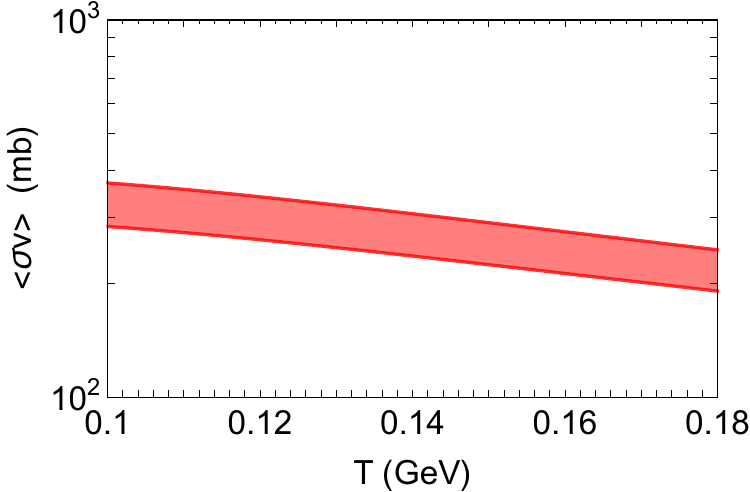}
			\caption{Thermally-averaged cross section for the process $d\pi  \to NN^\prime\pi$.
            }
			\label{fig:MedTer3}
\end{figure}

From the vacuum cross section for the process $d\pi \to NN'\pi$ in Eq.~(\ref{eq: medias_termicas}), we compute the corresponding thermally-averaged cross section presented in Fig.~\ref{fig:MedTer3}. The result exhibits a non-negligible magnitude and a mild decreasing trend across the temperature range relevant for a hadronic medium. 
This behavior suggests a smooth energy variation over the entire energy range of the underlying cross section where the thermal distribution is significant (for the temperatures of interest).

We now incorporate these results into a rate equation to model the time evolution of the deuteron multiplicity during the hadronic phase of heavy-ion collisions.

\section{The deuteron multiplicity}
\label{Abundance}

\subsection{The kinetic equation and hydrodynamic expansion}

To evaluate the time evolution of the deuteron yield, we employ an integro-differential equation successfully applied to various hadronization problems~\cite{xint-14-16,quase_livre,jpsi-18,dstar22,tcc22,zczcs,chi4274,psi2s}. Specifically, we adapt the framework detailed in the appendix of Ref.~\cite{quase_livre}, which leads to the following expression:
\begin{equation}\label{eq:rate_equation}
		\frac{dN_d(\tau)}{d\tau}= \langle \sigma_{d\pi \rightarrow NN^{\prime}\pi}\, v_{d\pi}\rangle \times n_\pi^{(eq)}(\tau) \times[N_d^{(eq)}(\tau)-N_d(\tau)].
	\end{equation}
where $N_d(\tau)$ denotes the abundance of the deuteron at proper time $\tau$, $n_\pi^{(eq)}(\tau)$ is the number density of pions forming the light hadron gas, assumed to be in thermal equilibrium and following the Boltzmann distribution  density 
\begin{align}
    n_i^{(eq)}(\tau) = \frac{1}{2\pi^2}\gamma_i g_im_i^2 T(\tau) K_2\left( \frac{m_i}{T(\tau)}\right),
    \label{statistical}
\end{align}
with $\gamma_i$, $g_i$, and $m_i$ being the fugacity, degeneracy and mass of the particle $i$, respectively, and $T(\tau)$ is the time-dependent temperature. The multiplicity, $N_i(\tau)$, is obtained by multiplying the abundance, $n_i(\tau)$,  by the volume $V(\tau)$.  In this sense,  $N_d^{(eq)}(\tau)$ is the number of deuterons in thermal equilibrium according to Eq.~(\ref{statistical}).


To model the hadron gas evolution, we employ the boost invariant Bjorken picture with an accelerated transverse expansion, in which the $\tau$-dependent temperature and volume are~\cite{xint-14-16,jpsi-18,dstar22,tcc22,zczcs,chi4274,psi2s}
\begin{align}
V(\tau) & = \pi \left[ R_C + v_C \left(\tau - \tau_C\right) + 
\frac{a_C}{2} \left(\tau - \tau_C\right)^2 \right]^2 \tau c , \nonumber \\
T(\tau) & =  T_C - \left( T_H - T_F \right) \left( \frac{\tau - 
\tau _H }{\tau _F - \tau _H}\right)^{\alpha}.
\label{eq: TempVol}
\end{align}
where $R_C$, $\upsilon_C$, $a_C$, and $T_C$ represent the transverse size, transverse velocity, transverse acceleration, and temperature at the critical time $\tau_C$, respectively. $T_H$ is the hadronization temperature at time $\tau_H$, and $T_F$ is the temperature at kinetic freeze-out time ($\tau_F$). These parameters are fixed for a hadronic medium formed in central Pb-Pb collisions with $\sqrt{s_{NN}} = 5.02$ TeV, according to \cite{ExHIC:2017smd, magalhaes_2024}, and are presented in Table \ref{tabela1}.

\begin{table}[!htbp]
       \caption{In the first three rows, we list the parameters used in Eq.~(\ref{eq: TempVol}) for central Pb–Pb collisions with $\sqrt{s_{NN}} = 5.02$ TeV. In the last row, we list the pion meson multiplicity, and $\alpha$ is the frequency used in the coalescence model.}
\centering \begin{tabular}{ccc}
\hline \hline $v_C(\mathrm{c})$ & $a_C\left(\mathrm{c}^2 / \mathrm{fm}\right)$ & $R_C(\mathrm{fm})$ \\
0.5 & 0.09 & 11 \\
\hline $\tau_C(\mathrm{fm} / \mathrm{c}) $&$ \tau_H(\mathrm{fm} / \mathrm{c}) $&$ \tau_F(\mathrm{fm} / \mathrm{c})$ \\
7.1 & 10.2 & 21.5 \\
\hline $T_C(\mathrm{MeV}) $&$ T_H(\mathrm{MeV}) $&$ T_F(\mathrm{MeV})$ \\
156 & 156 & 115 \\
\hline $N_\pi $&$\omega_d[\mathrm{MeV}]  $& $N_p = N_n$\\
713 &13.2  &  32 \\
\hline \hline \label{tabela1}
\end{tabular}
    \end{table}


\subsection{Initial conditions}

Now we define the initial conditions for the rate equation~(\ref{eq:rate_equation}). 
In Table~\ref{tabela1} we show  the initial multiplicity of pions. The  fugacity in Eq.~(\ref{statistical}) is a normalization parameter to adjust this initial multiplicity.



In the case of the deuteron, we test two possibilities. The first is given by the SHM, i.e. the use of Eq.~(\ref{statistical}). As in~\cite{ExHIC:2017smd}, we choose the baryon chemical potential  to be $\mu_B = 0$, which is appropriate for  a  Pb-Pb collision  at the energy of $\sqrt{s_{NN}} = 5.02$ TeV. Then, the deuteron fugacity becomes $\gamma_d=1$.  As a consequence, the initial condition for the deuteron according to the SHM is 
\begin{equation}
 N_{d}^{(SHM)} \equiv N_d^{(eq)}(\tau_H)=0.148.
 \label{init_cond_SHM}
\end{equation}

The second method for establishing the deuteron initial condition is the COM, which has the advantage of encoding the intrinsic structure of the system. This is achieved by calculating the hadronic multiplicity from the convolution of the constituent density matrix with the hadron Wigner function of the produced state. It is given by~\cite{ExHIC:2017smd}
\begin{align}
N^{(Coal)} (\tau_C) &  \approx 
g_{i} \prod_{j=1}^{n} \frac{N_j}{g_j} \,
\prod_{k=1}^{n-1} 
\frac{(4 \pi \sigma_k^2)^{\frac{3}{2}}}{V(\tau_C) \bigl(1 + 2 \mu_k T(\tau_C) \ \sigma_k^2 \bigr)} \nonumber \\ 
& \times 
\left[ \frac{4 \mu_k T(\tau_C) \ \sigma_k^2}{3 \bigl(1 + 2 \mu_k T(\tau_C) \  \sigma_k^2 \bigr)}
\right]^{l_i} ,
\label{dCoal}
\end{align}
In this expression, $g_j$ and $N_j$ are the degeneracy and number of the $j$-th constituent, and $\sigma_k = (\mu_k \omega)^{-1/2}$. The parameters $\mu_k$ (reduced mass) and $\omega  = 6 E_B$ (oscillator frequency) define the harmonic oscillator potential used to model the hadron internal structure. Since the deuteron is treated as a hadronic molecule in $S$-wave, the orbital angular momentum is $l_i=0$. 
Therefore, by employing the relevant constants and the parameters provided in Table \ref{tabela1} in Eq.~(\ref{dCoal}), we obtain
\begin{align}
    N_{d}^{(Coal)} \equiv N_{d}(\tau_H) =1.012.
    \label{initialconditionsCoal}
\end{align}

\subsection{Results and discussion}

We now compute the deuteron yield by using the thermally-averaged cross sections obtained previously into the rate equation~(\ref{eq:rate_equation}), applying the statistical and coalescence models for the deuteron initial condition, given by Eqs.~(\ref{init_cond_SHM}) and~(\ref{initialconditionsCoal}), respectively. 

\begin{figure}[!htbp]
			\center
				\includegraphics[scale=0.7]{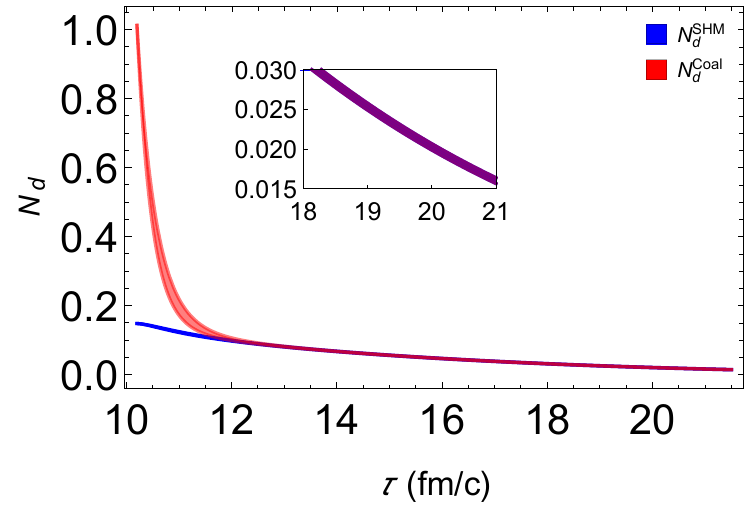}
				\caption{Deuteron multiplicity $N_d$ as a function of proper time $\tau$ in central Pb-Pb collisions at $\sqrt{s_{NN}} = 5.02~\mathrm{TeV}$. The curves correspond to initial conditions calculated from the statistical [Eq.~(\ref{init_cond_SHM})] and coalescence [Eq.~(\ref{initialconditionsCoal})] models. }
			\label{fig:compSHMCoal}
\end{figure}

Fig.~\ref{fig:compSHMCoal} shows the time evolution of the deuteron multiplicity throughout the duration of the hadronic phase in heavy-ion collisions. Firstly, it should be pointed out that our analysis reveals a sharp contrast between the initial conditions given by the statistical and coalescence models. Specifically, the COM predicts an initial deuteron abundance of about one order of magnitude higher than the SHM prediction. 

Interestingly, the evolution of the deuteron multiplicity further differentiates the models. The statistical model generates a final $N_d$ of the order of $\mathcal{O}(10^{-2})$ and exhibits a gradual decrease throughout the hadronic phase. On the other hand, the evolution of $N_d$ from the coalescence model undergoes rapid depletion, showing a stabilization for $\tau \gtrsim 14$ fm/c.  The statistical model retains a small fraction of its initial yield, with a survival rate of approximately $10\%$. The coalescence model, despite its large initial abundance, suffers even more severe attenuation, ending with less than $5\%$ of its initial population.

Most importantly, the main feature of these results is the convergence of the predictions of these models to a common final multiplicity value at the end of the hadronic phase: $N_d (\tau_F) \approx 0.015$. This convergence suggests that the final observable deuteron yield is insensitive to the initial production mechanism, potentially indicating a robust, model-independent final yield governed by the prevailing hadronic medium conditions at kinetic freeze-out.

\section{Concluding remarks}
\label{Conclusions}

In this work, we investigate the interactions of deuterons with the hadronic medium (assumed to be made of 
pions) formed in ultra-relativistic heavy-ion collisions. Our primary goal is to determine the thermally-averaged cross section for the deuteron dissociation  ($d\pi \rightarrow NN'\pi$) and to study its impact on the deuteron temporal  evolution. We calculate the cross sections  within the quasi-free approximation, where the $d\pi$ interaction is treated as the sum of pion scatterings off individual constituent nucleons. The underlying elementary $N\pi$ amplitudes are described by a hybrid effective model, combining the non-resonant background inspired by $\chi$PT with resonant contributions via Breit-Wigner parameterizations. This approach allows us to compute the crucial vacuum cross section, which is  then thermally averaged to incorporate the effects of the hot medium.

Our results yield two significant sets of findings. First, the cross section for deuteron absorption exhibits a substantial magnitude and a mild, decreasing trend with temperature. This behavior indicates that deuteron-pion interactions are a non-negligible channel throughout the hadronic phase, with a smoothly varying strength that remains relevant across the entire temperature range of the medium.

Second, and more interesting, when these cross sections are incorporated into the dynamical evolution of the deuteron yield, we observe a remarkable phenomenon: despite wildly different initial conditions --—where the coalescence model predicts an initial abundance one order of magnitude larger than the statistical hadronization model--- and drastically different evolutionary paths, both models converge to the same final deuteron multiplicity by the end of the hadronic phase.

This convergence to a unique final yield, $N_d (\tau_F) \approx 0.015$, is the central result of our analysis. It suggests that the final observable deuteron yield is independent of its initial production mechanism. Instead, it is  governed by the scattering processes within the hadronic medium and is ultimately determined by the thermodynamic conditions at kinetic freeze-out. 
It should also be mentioned that this finding aligns with previous analyses, such as that in Ref.~\cite{Staudenmaier:2021lrg}, where deuteron formation in HICs was simulated by implementing stochastic multi-particle reactions within a hadronic transport framework. That work similarly concluded that the final deuteron yield is practically independent of the assumed initial multiplicity.

Our work therefore provides a prediction that the deuteron yield observed in experiments may be a reliable probe of the late-stage hadronic environment as its final value is resilient to the theoretical uncertainties surrounding its formation.


\begin{acknowledgements}

This work was partly supported by the Brazilian agencies CNPq (L.M.A.: Grants No. 400215/2022-5, 308299/2023-0, 402942/2024-8)), CNPq/FAPERJ under the Project INCT-F\'isica Nuclear e Aplica\c{c}\~oes (Contract No. 464898/2014-5), FAPESB, and FAPESP Thematic
grants 2017/05660-0, 2019/07767-1, and 2020/04867-2.

\end{acknowledgements} 



\end{document}